# Tunable Dual-band IFA Antenna using *LC* Resonators

Nan Ni, Albert Humirang Cardona
AGILE RF INC.
SANTA BARBARA, CA, 93111 USA
nni@agilerf.com

*Abstract*— A tunable dual-band inverted F antenna (IFA) is presented in this paper. By placing a LC resonator on the radiating arm, dual-band characteristic is achieved. Especially, the capacitor in the resonator is a tunable thin-film BST capacitor, which has a 3.3:1 tuning ratio. The capacitance of the BST capacitors can be tuned by an external DC bias voltage. By varying the capacitance, both the lower band and the upper band of the IFA antenna can be tuned. And the total bandwidth can cover six systems, i.e., GSM-850, GSM-900, GPS, DCS, PCS, and UMTS.

## I. INTRODUCTION

In wireless communications, different systems used in different geographical regions have different frequency bandwidths. For example, In Europe, the GSM-900 standard has frequency bands of 890-915MHz and 935-960MHz for the uplink and downlink, respectively, The GSM-1800 (also called DCS-1800) uses 1710–1785 MHz and 1805–1880 MHz for the uplink and downlink, respectively. In North America, GSM-850 uses 824–849MHz for the uplink and 869–894MHz for the downlink. And GSM-1900 (also called PCS-1900) uses 1850–1910MHz for the uplink and 1930–1990MHz for the downlink, respectively. For 3G wireless systems, the UMTS in Europe uses 1900-1980 MHz, 2010-2025 MHz, and 2110-2170 MHz bands for terrestrial transmission. In North America, the CDMA200 use 824-849MHz, 869–894MHz, 1850-1910MHz, and 1930–1990MHz, respectively. Hence, for the portable cellular phones to be compatible with the various systems, dual-band antennas are needed [1]-[8].

Previous dual-band designs have been generally based on the dual-feed approach where the lower band and upper band antennas are put together into a compact structure [9] [10]. However, all these antennas will interfere with each other while transmitting or receiving the signals.

Another approach incorporated a *LC* resonator on a PIFA antenna to realize the dual-band characteristic [11]. This approach only has one signal feed. Therefore, there is no interference between the radiation elements. The *LC* resonator used is composed of a fixed capacitor and a fixed inductor. So the PIFA antenna only has fixed lower band and fixed upper band.

This paper presents a tunable dual-band Inverted-F antenna (IFA) which can cover a broader bandwidth compared to other dual-band antennas. The dual-band characteristic is realized by employing a *LC* resonator approach like in [11]. However, the capacitor used in the resonator is a tunable capacitor based on the thin-film BST technology. The capacitance of the BST capacitors can be tuned by an external DC voltage. By varying the capacitance, both the lower band and the upper band of the IFA antenna can be tuned to cover a broader bandwidth. Experiments show that totally six systems, i.e., GSM-850, GSM-900, GPS, DCS, PCS and UMTS, can be covered by using a 3.3:1 tunable capacitor. In addition, the planar structure of the IFA antenna makes it easy to incorporate it into cell phones or other wireless devices.

## II. ANTENNA ANALYSIS AND DESIGN

An IFA incorporating a *LC* resonator is illustrated in Fig. 1. It can be seen from Fig. 1(a) that the present antenna uses a single-feed approach. The grey area on the antenna represents the parallel *LC* resonator which is used to realize the dual-band characteristic. Fig. 1(b) shows the detailed structure of the *LC* resonator. The *LC* resonator is composed of an inductor L1, a tunable capacitor C1 and a DC block capacitor C2. The C2 is used to block the DC bias voltage so C1 will not be shorted. R1 is the DC bias resistor and is used to feed the DC bias voltage to C1.

The capacitor C1 used in the resonator is a Barium Strontium Titanate (BST) capacitor. BST generally has a high dielectric constant so that large capacitances can be realized in a relatively small area. Furthermore, BST has a permittivity that depends on the applied electric field. As a result, its capacitance can be tuned by changing a DC bias voltage across the BST capacitor thus voltage-controlled tunable capacitors can be produced. In addition, the bias voltage of the BST capacitor can be applied in either direction across a BST capacitor since the film permittivity is generally symmetric about zero bias. That is, BST dielectric does not exhibit a preferred direction for the electric field. One further advantage is that the electrical currents that flow through BST capacitors are relatively small compared to other types of semiconductor varactors.

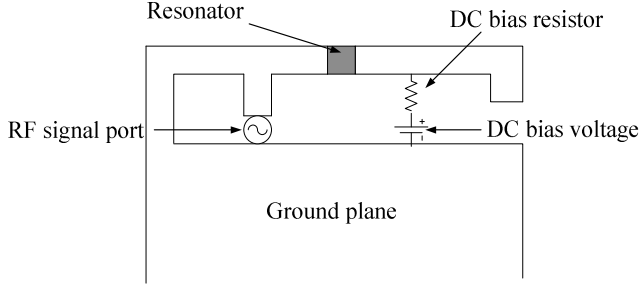

(a)

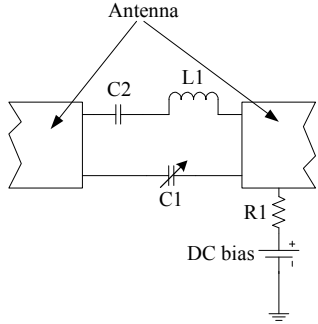

(b)

Fig. 1. (a) schematic form of the IFA antenna with *LC* resonator; (b) *LC* resonator, C1 is the tunable BST capacitor, L1 is a fixed inductor, C2 is a DC block capacitor, R1 is the DC bias resistor.

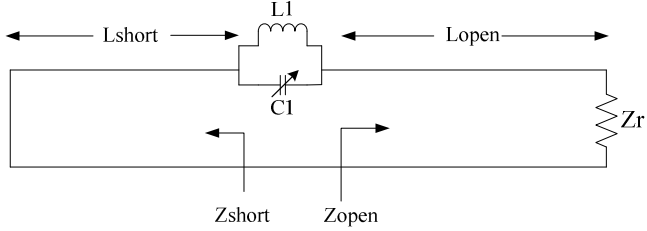

Fig. 2. Approximate transmission-line model.

Approximate analysis of the IFA with *LC* resonator can be obtained by utilizing a transmission-line model. The IFA can be taken as two transmission lines of characteristic impedance $Z_0$ with electrical lengths $L_{open}$ and $L_{short}$ connecting the open end to the resonator and the shorted end to the resonator, respectively, as illustrated in Fig. 2. Here, C2 is omitted since the characteristic of resonator is mainly decided by L1 and C1. The open end can be modeled as a load $Z_r$ while the shorted end can be taken as a shorted transmission line. The $L_{open}$ and $L_{short}$ are decided by the physical dimensions of the antenna. The parallel *LC* resonator can be thought of as being equivalent to an electrical length $L_{LC}(f)$ that depends on frequency. At frequency above its resonant frequency, the resonator becomes capacitive and effectively decreases the electrical length of the antenna. At frequency below its resonant frequency, the resonator becomes inductive and effectively increases the electrical length of the antenna. Adding the lengths $L_{open}$, $L_{short}$ and $L_{LC}(f)$, resonance of the antenna can be expressed as [11]

$$L_{open} + L_{short} + L_{LC}(f) = \lambda/4. \quad (1)$$

Since $L_{LC}(f)$ can have both negative and positive effective electrical lengths, dual resonance can be achieved.

Determining the values of L1 and C1 for producing dual resonance can be found by considering the impedance along the IFA. For one fixed value of C1 and one fixed value of L1, the impedance $Z_{lc}(f)$ of resonator is

$$Z_{lc}(f) = (\frac{1}{j\omega L} + j\omega C)^{-1}. \quad (2)$$

By defining $Z_{open}(f)$ and $Z_{short}(f)$ as the impedances seen from the *LC* resonator looking into the open and shorted end, respectively, at frequency *f*, we can write

$$Z_{Lc}(f) = Z_{short}(f) - Z_{open}(f). \quad (3)$$

as the condition for resonance at both $f_1$ and $f_2$. Solving the above equation at $f_1$ and $f_2$, we arrive at the following expressions for L and C [11]:

$$L = \frac{-3j}{4\pi f_1} \frac{Z_{LC}(f_2) Z_{LC}(f_1)}{2Z_{LC}(f_2) - Z_{LC}(f_1)}.$$

$$C = \frac{-j}{2\pi f_1}(\frac{1}{Z_{LC}(f_1)} - \frac{1}{j2\pi f_1 L}). \quad (4)$$

From the above analysis, it can be seen that the function of the *LC* resonator is to equate the impedance of antenna bodies at both ends of the resonator. Please note that L1 and C1 have fixed values in the above equations. If one of these two components, say C1, is tunable, then the equation will hold for two other frequencies, meaning the antenna could have tunable dual-frequency characteristic. In the present design, the capacitor C1 is a tunable BST capacitor and its capacitance could be controlled in a certain range by applying a DC voltage. For one capacitance of C1 and one inductance of L1, the resonator could equate the impedance of antenna bodies at two different frequencies. If the capacitance of C1 is changed, the resonator would equate the impedance of antenna bodies at two other frequencies. Therefore, the tunable dual-band characteristic can be realized.

### III. RESULTS

Fig. 3 shows the manufactured tunable dual-band IFA antenna. The dimension of the antenna arm is 10.6 x 49mm². The ground plane has the size of 80 x 50mm². The substrate is FR4 with thickness of 0.8mm.

A resonator including a BST tunable capacitor is soldered on the antenna arm. The cable is to feed the RF signal. The DC bias feed is on the back side of the ground plane. The component values are shown in TABLE I.

Fig. 4 shows the total bandwidth covered by the lower band and the upper band. Here, we use -6dB as the criterion for return loss, S11. The dotted line is the return loss under 0V i.e., when the BST capacitor has the biggest capacitance. The solid line is the return loss under 15V, i.e., when the BST capacitor has the smallest capacitance. It can be seen that the return loss is moving up along the frequency with increasing DC bias voltage.

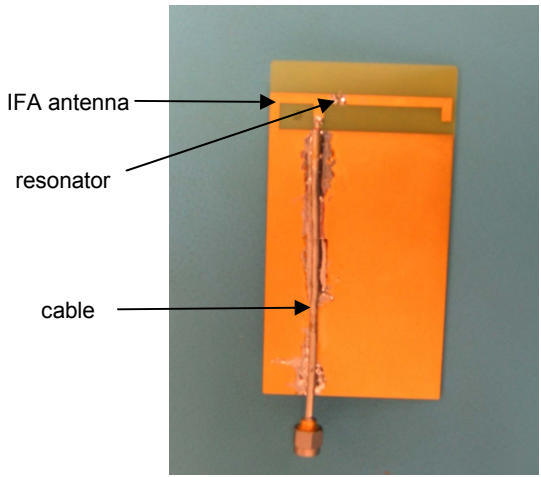

Fig. 3. Manufactured tunable dual-band IFA antenna with resonator on its arm.

TABLE I. COMPONENT VALUES IN THE RESONATOR

|  | C1 | C2 | L1 |
|---|---|---|---|
| **Values** | 2pF BST cap with 3.3:1 tunability under 15V DC bias voltage. | 68pF | 9.1nH |

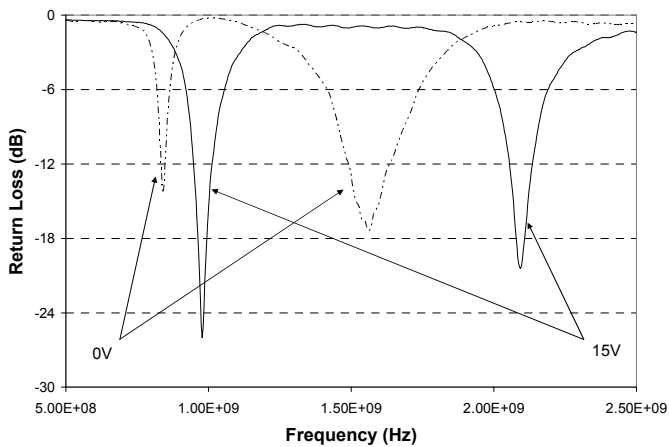

Fig. 4. The total bandwidth covered by the lower band and the upper band. The dotted line is the return loss under 0V i.e., when the BST capacitor has the biggest capacitance. The solid line is the return loss under 15V, i.e., when the BST capacitor has the smallest capacitance.

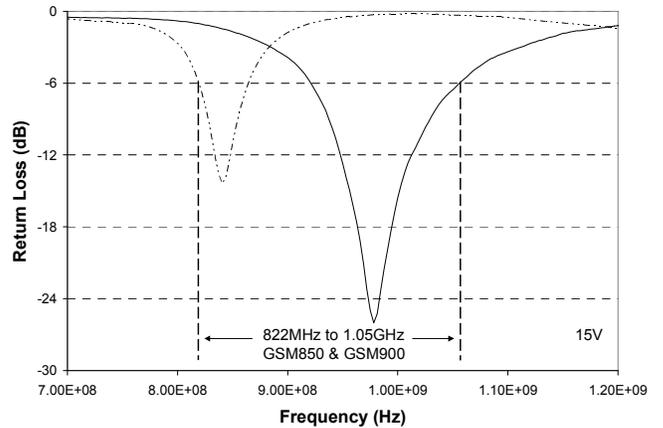

Fig. 5. The total bandwidth covered by the lower band. For a DC bias voltage from 0V to 15V, the total bandwidth is between 822MHz and 1.05GHz, which is enough to cover the GSM850 and GSM900 systems.

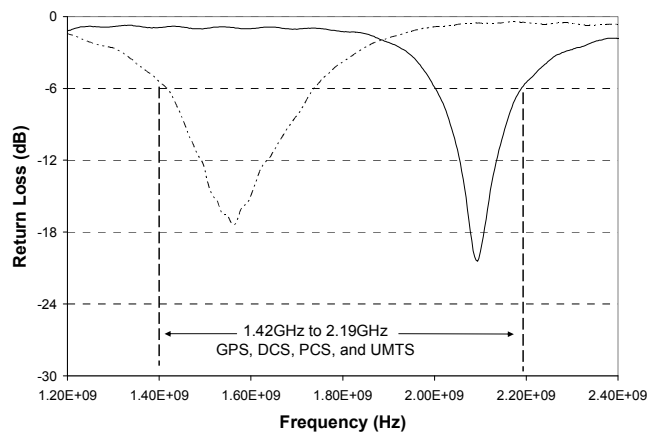

Fig. 6. The total bandwidth covered by the upper band. For a DC bias voltage from 0V to 15V, the total bandwidth is between 1.42GHz and 2.19GHz, which is enough to cover GPS, DCS, PCS, and UMTS systems.

Fig. 5 shows the lower band with different DC voltages, i.e., different capacitance of the BST tunable capacitor. The dotted line represents the return loss when applying 0V DC voltage, or in other words, when the BST tunable capacitor has its biggest value. Its bandwidth is between from 822MHz to 866MHz. The solid line represents the return loss when applying 15V DC voltage, or in other words, when the BST tunable capacitor has its smallest value. Its bandwidth is between 922MHz to 1.05GHz. Any other DC voltage (or any other capacitance) will have the bandwidth in between these two bandwidths. Therefore, the lower band is tunable by applying different DC bias voltages and its total bandwidth is between 822MHz and 1.05GHz, which is enough to cover the GSM850 and GSM900 systems.

Fig. 6 shows the upper band with different DC voltages, i.e., different capacitance of the BST tunable capacitor. The dotted line represents the return loss when applying 0V DC voltage, or in other words, when the BST tunable capacitor has its biggest value. Its bandwidth is between from 1.42GHz to 1.73GHz. The solid line represents the return loss when applying 15V DC voltage, or in other words, when the BST tunable capacitor has its smallest value. Its bandwidth is between 2.00GHz to 2.19GHz. Any other DC voltage (or any other capacitance) will have the bandwidth in between these two bandwidths. Therefore, the upper band is tunable by applying different DC bias voltages and its total bandwidth is between 1.42GHz and 2.19GHz, which is enough to cover the GPS, DCS, PCS, and UMTS systems.

From Fig. 4 to Fig. 6, it clearly can be seen that the present antenna show the tunable dual-band characteristic as expected. And its bandwidth can cover totally six different systems.

## IV. CONCLUSION

A tunable dual-band Inverted-F antenna (IFA) is presented in this paper. The tunable dual-band characteristic is realized by placing a *LC* resonator on the radiating arm. The *LC* resonator is comprised of a tunable BST capacitor C1, a DC blocking capacitor C2 and an inductor L1. A DC bias voltage is applied to the BST capacitor C1 in order to tune the capacitance. The IFA exhibits dual band characteristics, and both its lower band and upper band can be turned by tuning the capacitance of the BST capacitor C1. Measured data show that this tunable dual-band antenna can cover totally six different systems, i.e., GSM850, GSM900, GPS, DCS, PCS, and UMTS.

## REFERENCES

[1] Z. Wang, S. Kamalraj, C. C. Chiau, X. Chen, B. S. Collins, and S. P. Kingsley, "Dual-band Dielectric Antenna for WLAN Applications", *IEEE* International Workshop on Antenna Technologies (iWAT 2007), Cambridge, UKA, 21-23 Mach 2007.

[2] K.P. Ray and G. Kumar, "Tunable and dual band circular microstrip antenna with stubs", *IEEE Trans Antennas Propagat.*, vol. AP - 48, pp. 1036-1039, July 2000.

[3] B. Niu, O. Simeone, O. Somekh and A. M. Haimovich, "Ergodic and Outage Sum-Rate of Fading Broadcast Channels with 1-Bit Feedback," *IEEE Transactions on Vehicular Technology,* vol. 59, no. 3, pp. 1282 - 1293, Mar. 2010.

[4] Viratelle, D. and Langley, R.J.: "Dual band printed antenna for mobile telephone antenna applications", *IEE Proc., Microwaves, Antennas and Propagation*, 2000, 147(5), pp.381-384.

[5] B. Niu and A. M. Haimovich, "Interference Subspace Tracking for Network Interference Alignment in Cellular Systems," *IEEE Global Communications Conference (Globecom)*, Honolulu, November 2009, pp. 1 - 5.

[6] K. M. Luk, C. H. Lai and K. F. Lee, "Wideband L-probe-feed patch antenna with dual-band operation for GSM/PCS base stations", *Electronics Letters*, July 1999, vol.35, pp.1123-1124.

[7] B. Niu, O. Simeone, O. Somekh and A. M. Haimovich, "Throughput of Two-Hop Wireless Networks with Relay Cooperation," *45th Annual Allerton Conference on Communication, Control, and Computing*, Monticello, IL, September 2007.

[8] R.A. Fenner and E.J. Rothwell, "Bandwidth Extension of a Multi-Band Body Worn Antenna Vest," URSI North American Radio Science Meeting, Ottawa, Canada, July 23-26, 2007.

[9] P. S. Hall, Z. D. Liu, and D. Wake, "Dual frequency planar inverted-f antenna," *IEEE Trans. Antennas Propagat.*, vol. 45, pp. 1451–1458, Oct. 1997.

[10] C. R. Rowell and R. D. Murch, "A compact PIFA suitable for dualfrequency 900/1800 MHz operation," *IEEE Trans. Antennas Propagat.*, vol. 46, pp. 596–598, Apr. 1998.

[11] G. K. H. Lui and R. D. Murch, "Compact dual-frequency PIFA designs using LC resonators," *IEEE Trans. Antennas Propagat.*, vol. 49, pp. 1016–1019, July 2001.